\documentclass{article}%
\pdfoutput=1
\usepackage{amsmath}
\usepackage{amsfonts}
\usepackage{amssymb}
\usepackage{graphicx}
\usepackage{epstopdf}
\usepackage{cite}
\usepackage[usenames]{color}%
\providecommand{\U}[1]{\protect\rule{.1in}{.1in}}
\newcommand{\ba}{\begin{array}}
\newcommand{\ea}{\end{array}}
\newcommand{\Dsl}[1] { \setbox0=\hbox{$#1$}     
\dimen0=\wd0   \setbox1=\hbox{/} \dimen1=\wd1  \ifdim\dimen0>\dimen1        
 \rlap{\hbox to \dimen0{\hfil/\hfil}}  #1 \else \rlap{\hbox to \dimen1{\hfil$#1$\hfil}}  /  \fi  }
\newcommand{\bea}{\begin{eqnarray}}
\newcommand{\eea}{\end{eqnarray}}

\newcommand{\ns}{\Dsl{n}}
\newcommand {\nbs}{\Dsl{\bar n}}

\begin{document}
\noindent

\title {  \Large $\chi_{cJ}\rightarrow e^{+}e^{-}$ decays revisited }


\author{ N. Kivel \thanks{On leave of absence from St.~Petersburg Nuclear Physics
Institute, 188350, Gatchina, Russia} \  
and  M. Vanderhaeghen 
\\[3mm]
{\it Helmholtz Institut Mainz, Johannes Gutenberg-Universit\"at, D-55099
Mainz, Germany}
 \\
{\it Institut f\"ur Kernphysik, Johannes
Gutenberg-Universit\"at, D-55099 Mainz, Germany } 
 }

\date{}

\maketitle

\vspace*{1cm}

\begin{abstract}
We present  a calculation of the width  for  $\chi_{cJ}\rightarrow e^{+}e^{-}$ decay.  
The amplitude of the process is computed  within the NRQCD framework. 
The leading-order contribution is described by  two terms  associated  with
the two different integration domains  in the electromagnetic  loop describing two-photon 
 annihilation of the heavy  quark-antiquark pair. The corresponding  operators are defined in the framework 
 of  NRQCD. 
 The matrix element of  one of these operators describes a configuration  with an ultrasoft photon 
 and can be associated with the higher Fock  state of the heavy meson.
 In order to compute this contribution  we use the heavy hadron chiral perturbation theory.  We obtain that  
 this contribution is numerically dominant.  The obtained estimates for the decay widths of the $\chi_{c1}$ and $\chi_{c2}$ states are 
 $0.09$ eV and $0.06$ eV, respectively. 
 \end{abstract}

\newpage

\section{Introduction}
\label{int}

The leptonic decays of C-even charmonium states into a lepton
pair have a very small branching ratio because they can only occur via a
two-photon intermediate state $\chi_{cJ}\rightarrow\gamma^{*}\gamma
^{*}\rightarrow e^{+}e^{-}$. However with the high-luminosity BEPC-II
$e^{+}e^{-}$-accelerator operating on the charmonium energy region 
such measurements of direct production cross section $e^{+}e^{-}\rightarrow
\chi_{cJ}$ become feasible. The study of the  mechanism of the production of $C$-even quarkonium states is
especially interesting in view of the  production of higher resonances such as
the exotic charmonium-like state X(3872).

The decays  $\chi_{cJ}\rightarrow l^{+}l^{-}$  have already been studied
 long time ago in Ref.\cite{Kuhn:1979bb}. The authors considered
the  quarkonium description  and a vector dominance model (VDM) in order to 
describe  the decay amplitudes of charmonium states with $J=1,2$. 
It was found that a naive  quarkonium description is problematic because  of 
 infrared logarithmic divergencies arising   in the integrals 
describing the quark-photon annihilation loop. Such divergencies indicate that the corresponding contribution   is also sensitive to
 long distance physics. The corresponding integrals in Ref.\cite{Kuhn:1979bb} have been regularized by introducing
the binding energy $M_{\chi}-2m_{Q}\simeq$ $500$ MeV.  The  
 numerical estimates obtained in this way give very small  value of the widths,  smaller thаn 
 bounds derived from  analyticity and unitarity  in the same work.   Probably, a more
realistic estimate was obtained using  generalized VDM  which yields a larger numerical value for the widths,
  consistent
with the unitarity constrains: $\Gamma\lbrack\chi_{c1}\rightarrow e^{+}%
e^{-}]\simeq0.46$~eV and $\Gamma\lbrack\chi_{c2}\rightarrow e^{+}e^{-}%
]\simeq0.014$~eV \cite{Kuhn:1979bb}. 

Recently the decay rate of the $\chi_{c1}$ state
was again estimated in \cite{Denig:2014fha} using the VDM approach, with result
$\Gamma\lbrack\chi_{c1}\rightarrow e^{+}e^{-}]\simeq0.1$~eV.  The short
distance contributions describing a  configuration with 
highly virtual photons were considered in this framework as unknown contact vertices
giving rise to a theoretical uncertainty.

The aim of the present  work is to provide a more systematic description of the decay
amplitudes for $\chi_{cJ}\rightarrow\gamma^{*}\gamma^{*}\rightarrow e^{+}
e^{-}$ process using the NRQCD factorization framework 
\cite{Bodwin:1994jh, Beneke:1997zp}, see also review \cite{Brambilla:2004jw} and references theirin.  
This technique  allows one to perform a systematic
description of heavy quarkonium states using the small relative velocity of heavy quarks 
and the small QCD running coupling at short distances.  Within this framework
 we  associate the IR-divergencies found in \cite{Kuhn:1979bb}
with a specific quark-photon operator which describes a configuration with an ultrasoft 
 photon. 
 The matrix element of this operator  describes the overlapping  with  the higher Fock  state  which
 consists of heavy quark-antiquark  and photon. 
 This allows us to perform a systematic  separation and description of the
different contributions relevant in a  leading-order expansion in small velocity $v$. 

Our paper is organized as follows.  In Sec.~\ref{kin} we briefly describe our notation
and provide  definitions of various  quantities used in the following.  Sec.~\ref{fact} is devoted to  the
investigation of the one-loop integral which describes the leading-order contribution. 
In this  section we establish the dominant regions and provide a description of  the amplitude 
within the NRQCD factorization framework.  Sec.~\ref{phen} is devoted to the calculation of 
the ultrasoft photon matrix element in the heavy hadron chiral effective theory (HH$\chi$PT).
Furthermore, a estimate  of the decay rates is given. 
In Sec.~\ref{conc}  we briefly  summarize our results.   

\section{Kinematics and notation}

\label{kin}

Let us start from the description of the decay kinematics $\chi_{cJ}(P)\rightarrow
e^{+}(l_{1})e^{-}(l_{2})$.  The initial state momentum can be written as
\begin{equation}
P=M_{\chi}~\omega,~~\omega^{2}=1,
\end{equation}
where $\omega$ denotes the charmonium velocity. In the following,  we  consider
the charmonium rest frame  which implies%
\begin{equation}
\omega=(1,\vec{0}). \label{RF}%
\end{equation}
 The small relative velocity of heavy quarks in the bound state is denoted as
$v$.  Neglecting lepton masses, the lepton momenta can be written as
\begin{equation}
l_{1}=M_{\chi}\frac{n}{2},~\ l_{2}=M_{\chi}\frac{\bar{n}}{2},~
\end{equation}
where $n$ and $\bar{n}$ denote the light-like vectors which satisfy
$(n\cdot\bar{n})=2$.  Any  4-vector $V^{\mu}$ can be expanded as
\begin{equation}
V^{\mu}=\left(  V\cdot n\right)  \frac{\bar{n}^{\mu}}{2}+\left(  V\cdot\bar
{n}\right)  \frac{n^{\mu}}{2}+V_{\bot}^{\mu},
\end{equation}
where $V_{\bot}$ denotes the components  transverse  to the
light-like vectors   : $\left(  V_{\bot}\cdot n\right)  =\left(  V_{\bot}
\cdot\bar{n}\right)  =0$.  Similarly, one can also write a decomposition
\begin{equation}
V^{\mu}=\left(  V\cdot\omega\right)  \omega^{\mu}+V_{\top}^{\mu},
\end{equation}
where $V_{\top}$ denotes the  component which is orthogonal to the
velocity $\omega$: $\left(  \omega\cdot V_{\top}\right)  =0$. In the following we assume that in the rest  frame
\begin{equation}
\omega=\frac{n}{2}+\frac{\bar{n}}{2}.
\end{equation}

The momenta of the heavy quark and antiquark  with mass $m$ which form a quarkonium state can be  written  as
\begin{equation}
p_{1}=\frac{1}{2}P+\Delta,~~p_{2}=\frac{1}{2}P-\Delta, \label{pi:def}%
\end{equation}
where the relative momentum $\Delta$ satisfies%
\begin{equation}
\left(  \Delta\cdot\omega\right)  =0,~\ \Delta^{2}=-\vec{\Delta}^{2}.
\end{equation}
The heavy quarks which create a bound state are non-relativistic, implying
that the relative velocity $v\sim\Delta/m$ is quite small: $v\ll 1$.

The power counting rules of NRQCD has been established in \cite{Lepage:1992tx,
Bodwin:1994jh}. Following these arguments we assume that the mass $m$ is large
enough and that  the most important scales such as mass $m$, typical three-momentum
of the heavy quark $\sim mv$ and its typical kinetic energy $\sim mv^{2}$
satisfy
\begin{equation}
\left(  mv^{2}\right)  ^{2}\ll(mv)^{2}\ll m^{2}\text{.}%
\end{equation}
Integrating out the modes with hard momenta $p_{h}\sim m_{c}$  one passes
onto the effective theory NRQCD which describes the modes with the soft momenta
$p_{s}\sim mv$. If the scale $mv\gg\Lambda_{QCD}$ then one can integrate over
the soft region together with potential gluons with momenta \cite{Pineda:1997bj,Pineda:1997ie,Beneke:1997zp, Brambilla:1999qa, Brambilla:1999xf}
\begin{equation}
\text{ \ }p_{0}\sim mv^{2},~\ \vec{p}\sim m\vec v,
\end{equation}
After this one obtains a new effective theory  which is known as potential NRQCD
(pNRQCD). For a more detailed information about these effective theories see
Ref.\cite{Brambilla:2004jw} and references therein.

The charm quark mass $m_{c}\simeq1.5$ GeV is not large enough compared
to $\Lambda_{QCD}$ therefore in this case one can only factorize the effects
at momentum scales of order $m_{c}$. However,  in the QED sector one can also
consider the possibility to integrate over the soft region too. As we will show further on,
 such situation is relevant for the factorization of the electromagnetic
loop describing the transition $c\bar{c}\rightarrow\gamma^{\ast}\gamma^{\ast
}\rightarrow e^{+}e^{-}$.

After factorization of the hard contribution, the  nonpertubative QCD dynamics is described by the matrix elements of
appropriate operators defined in an effective theory.  In the following, we will need the following  set of NRQCD operators which describe the matrix
elements  between the charmonium and the vacuum states:
\begin{equation}
\mathcal{O}^{\sigma}(^{3}S_{1})=\chi_{\omega}^{\dag}\gamma_{\top}^{\sigma}%
\psi_{\omega}, \label{O3S1}%
\end{equation}%
\begin{equation}
\mathcal{O}(^{3}P_{0})=-\frac{1}{\sqrt{3}}~\chi_{\omega}^{\dag}\left(
\frac{-i}{2}\right)  \overleftrightarrow{D}_{\top}^{\alpha}\gamma_{\top
}^{\alpha}\psi_{\omega}, \label{O3P0}%
\end{equation}%
\begin{equation}
\mathcal{O}^{\beta}(^{3}P_{1})=\frac{1}{2\sqrt{2}}~\chi_{\omega}^{\dag
}\overleftrightarrow{D}_{\top}^{\alpha}\left(  \frac{-i}{2}\right)  \left[
\gamma_{\top}^{\alpha},\gamma_{\top}^{\beta}\right]  \gamma_{5}\psi_{\omega},
\label{O3P1}%
\end{equation}%
\begin{equation}
\mathcal{O}^{\alpha\beta}(^{3}P_{2})=\chi_{\omega}^{\dag}\left(  \frac{-i}%
{2}\right)  \overleftrightarrow{D}_{\top}^{(\alpha}\gamma_{\top}^{\beta)}%
\psi_{\omega}, \label{O3P2}%
\end{equation}
where the covariant derivative $iD_{\mu}=i\partial_{\mu}+gA_{\mu}$,
\ $\overleftrightarrow{D}_{\top}=\overrightarrow{D}_{\top}-\overleftarrow{D}%
_{\top}$. Furthermore,  we use the covariant four-component fields $~\psi_{\omega}$%
,~$\chi_{\omega}$ to describe the soft quark and antiquarks within the
NRQCD framework. These fields satisfy%
\begin{equation}
\chi_{\omega}^{\dag}~\Dsl{\omega}=-\chi_{\omega}^{\dag},~\ \ \Dsl{\omega}%
\psi_{\omega}=\psi_{\omega}\text{.}%
\end{equation}
Using Eq.(\ref{RF}), one can show that the operators in Eqs.(\ref{O3S1})-(\ref{O3P2})
can be reduced to the set of  well-known non-relativistic operators constructed from two-component Pauli spinors.

The matrix elements of these  operators are well known in the literature and can be written as
\bea
\left\langle 0\right\vert \psi_{\omega}^{\dag}\gamma_{\top}^{\sigma}%
\chi_{\omega}\left\vert J/\psi\right\rangle =\epsilon_{\psi}^{\sigma
}~\left\langle \mathcal{O}(^{3}S_{1})\right\rangle,
 \label{me3S1}\\
 \left\langle 0\right\vert \psi_{\omega}^{\dag}\gamma_{\top}^{\sigma}%
\chi_{\omega}\left\vert \psi'\right\rangle =\epsilon_{\psi'}^{\sigma
}~\left\langle \mathcal{O}'(^{3}S_{1})\right\rangle ,
\label{me3S1psi}
\eea
\begin{equation}
\left\langle 0\right\vert 
\mathcal{O}(^{3}P_{0})
\left\vert \chi_{c0}\right\rangle =i\left\langle
\mathcal{O}(^{3}P_{0})\right\rangle , \label{me3P0}%
\end{equation}%
\begin{equation}
\left\langle 0\right\vert 
\mathcal{O}^{\sigma}(^{3}P_{1})\left\vert \chi_{c1}\right\rangle =i\epsilon_{\chi}^{\sigma}~\left\langle
\mathcal{O}(^{3}P_{0})\right\rangle , \label{me3P1}%
\end{equation}%
\begin{equation}
\left\langle 0\right\vert 
\mathcal{O}^{\alpha\beta}(^{3}P_{2})
\left\vert \chi_{c2}\right\rangle =i\epsilon_{\chi}^{\alpha\beta
}~\left\langle \mathcal{O}(^{3}P_{0})\right\rangle . \label{me3P2}%
\end{equation}
The constants $\left\langle \mathcal{O}(^{3}S_{1})\right\rangle $ and
$\left\langle \mathcal{O}(^{3}P_{0})\right\rangle $  are
related to the  value of the charmonium wave functions at the origin
\bea
\left\langle \mathcal{O}(^{3}S_{1})\right\rangle =\sqrt{2N_{c}} \sqrt{2M_{\psi}}%
\sqrt{\frac{1}{4\pi}}~R_{10}(0),
\label{def:R10}
\\
\left\langle \mathcal{O}'(^{3}S_{1})\right\rangle =\sqrt{2N_{c}} \sqrt{2M_{\psi'}}%
\sqrt{\frac{1}{4\pi}}~R_{20}(0),
\label{def:R20}
\eea%
\begin{equation}
\left\langle \mathcal{O}(^{3}P_{0})\right\rangle =\sqrt{2N_{c}}\sqrt
{2M_{\chi_{c0}}}\sqrt{\frac{3}{4\pi}}R_{21}^{\prime}(0), \label{def:R21}%
\end{equation}
where $R_{nl}(r)~$is the radial part of the Schr\"odinger wave function and $R_{nl}^{\prime
}(r)$ denotes its derivative. The \textit{rhs} of Eqs.(\ref{me3P0})-\ref{me3P2}) depends on of 
the same  constant $\left\langle \mathcal{O}%
(^{3}P_{0})\right\rangle $ due to the spin symmetry of the leading
non-relativistic action \cite{Bodwin:1994jh}. \ The polarization vectors
$\epsilon_{\psi}^{\sigma}(\lambda)$, $\epsilon_{\chi}^{\beta}(\lambda)$ and
$\epsilon_{\chi}^{\alpha\beta}(\lambda)$ correspond to  spin-1 and spin-2
charmonium states, respectively. They are normalized to satisfy
\begin{equation}
\sum_{\lambda}\epsilon_{X}^{\sigma}(\lambda)\left\{  \epsilon_{X}^{\rho
}(\lambda)\right\}  ^{\ast}=-g^{\sigma\rho}+\frac{P^{\sigma}P^{\rho}}{M^{2}_{X}%
},\
\end{equation}
with $X=\left\{  J/\psi,~\chi_{c1}\right\}  $ and%
\begin{equation}
\sum_{\lambda}\epsilon_{\chi}^{\alpha\beta}(\lambda)\left\{  \epsilon_{\chi
}^{\alpha^{\prime}\beta^{\prime}}(\lambda)\right\}  ^{\ast}=\frac{1}%
{2}M_{\alpha\alpha^{\prime}}M_{\beta\beta^{\prime}}+\frac{1}{2}M_{\alpha
\beta^{\prime}}M_{\beta\alpha^{\prime}}-\frac{1}{3}M_{\alpha\beta}%
M_{\alpha^{\prime}\beta^{\prime}},~\
\end{equation}
with $M_{\alpha\beta}=-g_{\alpha\beta}+P_{\alpha}P_{\beta}/M_{\chi_{c2}}^{2}$.

The decay amplitudes $\chi_{cJ}\rightarrow e^{+}e^{-}$ \ are \ defined as
\begin{equation}
\left\langle e^{+}e^{-};out\right\vert \left.  in;~\chi_{cJ}\right\rangle
= i(2\pi)^{4}\delta(l_{1}+l_{2}-P)~\mathcal{A}_{J},
\end{equation}
with%
\begin{equation}
\mathcal{A}_{J}=\bar{u}_{n}\Gamma_{J}v_{\bar{n}}~T_{J},~
\end{equation}
and where $\bar{u}_{n}$ and $\bar{v}_{\bar{n}}$ denotes the spinors of the
massles lepton and antilepton, respectively%
\begin{equation}
\bar{u}_{n}=\bar{u}(l_{1})\frac{\nbs\ns}{4},~\ v_{\bar{n}}=\frac{\nbs\ns}%
{4}v(l_{2}),
\end{equation}
and%
\begin{equation}
\Gamma_{1}=\epsilon_{\chi}^{\sigma}\gamma_{\bot \sigma}\gamma_{5}%
,~\Gamma_{2}=\epsilon_{\chi}^{\sigma\rho}n_{\rho}\gamma_{\bot\sigma}.\
\end{equation}

The leading-order contribution  to these amplitudes arises from the annihilation
of  heavy quarks into two photons which create the outgoing lepton pair, see Fig.~\ref{diagrams}.  
If the one-loop integral is dominated by the  hard region
where both photons and heavy quark are highly off-shell then one can expect
that such process can be described within the NRQCD approach and the amplitude
can be factorized into hard and soft parts. In the next section we consider
this possibility in more detail.

\section{Factorization of decay amplitudes in NRQCD}
\label{fact}

\begin{figure}[ptb]
\centering
\includegraphics[
height=1.5301in,
width=5.5251in]
{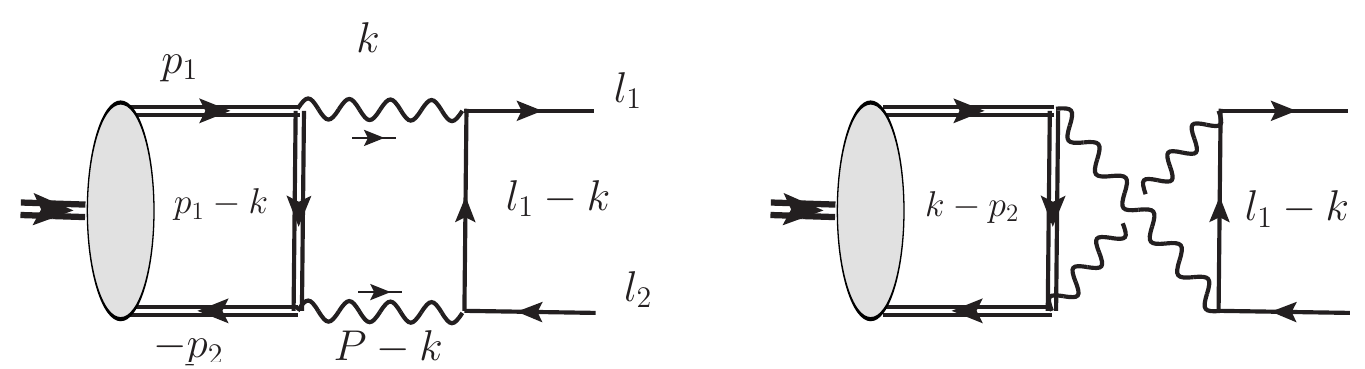}
\caption{One-loop diagrams describing the annihilation of $\chi_{cJ}$ 
into an $e^{+}e^{-}$ pair.}
\label{diagrams}
\end{figure}

The leading-order in $\alpha_{s}$ diagrams describing the  $e^{+}e^{-}$ decay of $C$-even
charmonia are shown in Fig.~\ref{diagrams}. These  one-loop diagrams are
 constructed from the photon, lepton and heavy quark (double lines).
 The diagrams  in Fig.~\ref{diagrams} 
can be  computed in the heavy quark mass limit $m\rightarrow\infty$, performing
an expansion in the small parameter $\Delta/m\sim v$.  
Let us to start  from a naive guess that the dominant  contribution 
is only provided by the hard region where  the loop momentum $k_{\mu}\sim m$, 
and therefore all propagators are far of off-shell.  The leading-order
contribution in $1/m$ is provided by projections onto the leading-order
operators $\mathcal{O}(^{3}P_{J})$ described in Eqs.(\ref{me3P0})-(\ref{me3P2}). 
The technical details are well known in the literature, see e.g.  \cite{Kuhn:1979bb}. 
The resulting  expressions can be presented as
\begin{equation}
\mathcal{A}_{1}=\epsilon_{\chi}^{\nu}~i\left\langle \mathcal{O}(^{3}%
P_{0})\right\rangle e^{2}\sqrt{2}\int dk~\frac{~\bar{u}_{n}\gamma_{\alpha
}(\Dsl{l}_{1}-\Dsl{k})\gamma_{\beta}v_{\bar{n}}}{\left[  \left(  k-l_{1}\right)
^{2}\right]  \left[  k^{2}\right]  \left[  \left(  k-P\right)  ^{2}\right]
}\frac{1}{4}\text{Tr}\left[  \mathcal{P}_{1\mu\nu}\Gamma^{\alpha\beta\mu
}(k)\right]  ,\ \label{Axee1}%
\end{equation}%
\begin{equation}
\mathcal{A}_{2}=\epsilon_{\chi \mu\nu}~i\left\langle \mathcal{O}(^{3}%
P_{0})\right\rangle e^{2}\int dk~
\frac{\bar{u}_{n}\gamma_{\alpha}%
(\Dsl{l}_{1}-\Dsl{k})\gamma_{\beta}v_{\bar{n}}}{\left[  \left(  k-l_{1}\right)
^{2}\right]  \left[  k^{2}\right]  \left[  \left(  k-P\right)  ^{2}\right]
}\frac{1}{4}\text{Tr}\left[  \mathcal{P}_{2}^{\nu}\Gamma^{\alpha\beta\mu
}(k)\right]  , \label{Axee2}%
\end{equation}
where the square brackets for the propagators denote the standard Feynman
prescription $[A]^{-1}\equiv\left[  A+i\varepsilon\right]  ^{-1}$.
The corresponding contribution to the amplitude $\chi_{c0}\rightarrow e^{+}e^{-}$ vanishes
 and therefore is suppressed by a power of $v$ and  will not be
considered it in this work. 

 The total structure  of the integrands in expressions  (\ref{Axee1}) and
(\ref{Axee2}) can be divided  into the lepton and heavy quark parts. 
The  lepton part has 
$\bar{u}_{n}\dots v_{\bar{n}}$ in the numerator  and  includes the photon and lepton propagators in the denominators.
 The heavy quark part  is given by Tr$\left[  \mathcal{P}_{i}\Gamma^{\alpha\beta\mu
}(k)\right]  $.   We introduced the projections $\mathcal{P}_{J}$ onto charmonium states
\begin{equation}
\mathcal{P}_{1}^{\mu\nu}=\frac{1}{4}(1+\Dsl{\omega})(\gamma_{\top}^{\mu}
\gamma_{\top}^{\nu}-\gamma_{\top}^{\nu}\gamma_{\top}^{\mu})\gamma_{5},
\end{equation}%
\begin{equation}
\mathcal{P}_{2}^{\nu}=(1+\Dsl{\omega})\gamma_{\top}^{\nu}.
\end{equation}
The expression for $\Gamma^{\alpha\beta\mu}(k)$ reads
\begin{equation}
\Gamma^{\alpha\beta\mu}(k)=\frac{1}{2m}\left\{  \gamma^{\mu}\hat{D}
_{Q}(k)+\hat{D}_{Q}(k)\gamma^{\mu}\right\}  +\hat{D}_{Q}^{\prime\mu
}(k),\label{G(k)}
\end{equation}
with
\begin{equation}
\hat D_{Q}=\frac{i(iee_{Q})^{2}}{\left[  k^{2}-2m(k\omega)-\vec{\Delta}^{2}\right]
}\left\{  \gamma^{\beta}(m\Dsl\omega-\Dsl k+m)\gamma^{\alpha}+\gamma^{\alpha}
(\Dsl k-m\Dsl\omega+m)\gamma^{\beta}\right\}  ,\label{DQ}
\end{equation}
\begin{align}
\hat D_{Q}^{\prime\mu} &  =\frac{i(iee_{Q})^{2}}{\left[  k^{2}-2m(k\omega
)-\vec{\Delta}^{2}\right]  }\left\{  \gamma^{\beta}\gamma^{\mu}\gamma^{\alpha
}+\gamma^{\alpha}\gamma^{\mu}\gamma^{\beta}\right\}  \nonumber\\
&  +\frac{i(iee_{Q})^{2}2k^{\mu}}{\left[  k^{2}-2m(k\omega)-\vec{\Delta}%
^{2}\right]  ^{2}}\left\{  \gamma^{\beta}(m\Dsl \omega-\Dsl k+m)\gamma^{\alpha}%
-\gamma^{\alpha}(\Dsl k-m\Dsl \omega+m)\gamma^{\beta}\right\}  ,\label{DQpr}%
\end{align}
where $e_{Q}$ is the charge of the heavy quark ($e_{c}=2/3$). The small squared
relative momentum  $\vec{\Delta}^{2}\sim(mv)^{2}$ which appears in the heavy quark
propagator provides an IR-regularization and  can be neglected if it is
not required. With this regularization the traces and loop integrals are computed in four dimensions
with $dk\equiv d^{4}k/(2\pi)^{4}$.

The expressions (\ref{Axee1}) and (\ref{Axee2}) have been obtained by
expanding the heavy quark fields in position space
\begin{equation}
c(y)\simeq e^{-im(\omega y)}\left[  1+y\cdot\partial+\frac{1}{2m}
i\Dsl D_{\top}\right]  \psi_{\omega}(0)
\end{equation}
and projecting the soft quark fields $\chi_{\omega}^{\dag}$ and $\psi_{\omega
}$ onto leading-order operators (\ref{O3P0})-(\ref{O3P2}). The terms $\sim y\cdot\partial
\psi_{\omega}~$ (arising  from the multipole expansion of the soft quark
field arguments ) lead to the expansion of the integrand with respect to small relative
momentum $\Delta$ giving the contribution $D_{Q}^{\prime\mu}$. The terms
proportional to $\sim\frac{1}{2m}\Dsl D_{\top}$ give the contribution with $\hat{D}_{Q}$.
 The evaluation  of the integrals in Eqs. (\ref{Axee1}) and (\ref{Axee2}) gives%
\begin{equation}
\mathcal{A}_{1}=\bar{u}_{n}\Gamma_{1}v_{n}~i\left\langle \mathcal{O}(^{3}%
P_{0})\right\rangle \frac{\alpha^{2}}{m^{3}}e_{Q}^{2}~2\sqrt{2}\ln\frac{m^{2}%
}{2\vec{\Delta}^{2}}, \label{A1}%
\end{equation}%
\begin{equation}
\mathcal{A}_{2}=\bar{u}_{n}\Gamma_{2}v_{n}~i\left\langle \mathcal{O}(^{3}%
P_{0})\right\rangle \frac{~\alpha^{2}}{m^{3}}e_{Q}^{2}2\left(  2\ln\vec{\Delta
}^{2}/m^{2}+\frac{2}{3}\left(  \ln2-1+i\pi\right)  \right)  . \label{A2}%
\end{equation}
These expressions are in agreement with the results obtained in
Ref. \cite{Kuhn:1979bb}. We obtain that both amplitudes depend on the \ large
logarithm $\sim\ln\vec{\Delta}^{2}/m^{2}$ which is sensitive to the soft scale
$\vec{\Delta}^{2}$. This shows that  the starting assumption about  one
dominant region $k\sim m$ is incorrect.  There must be at least one more
domain where  some propagators in the loop integral are soft. One can expect
that the additional region is associated with the configuration when one of
the photons is soft. In this case the propagator of the heavy quark is also
soft and the hard configuration is described by the tree level subdiagram
describing the annihilation $c\bar{c}\rightarrow e^{+}e^{-}$ through  one
photon.  

In order to get an idea about the explicit definition of this region it is
useful to investigate the integrals of diagrams in Fig.~\ref{diagrams} within
the threshold expansion technique worked out in Ref.\cite{Beneke:1997zp}.
According to this analysis the threshold kinematics is described by the
following regions
\begin{equation}
\text{hard}:\text{ }k_{\mu}\sim m, \label{HR}%
\end{equation}%
\begin{equation}
\text{soft}:k_{\mu}\sim mv,~ \label{SR}%
\end{equation}%
\begin{equation}
\text{potential}:k_{0}\sim mv^{2},~\ \vec{k}~\sim mv,\ \label{PR}%
\end{equation}%
\begin{equation}
\text{usoft}:~k_{\mu}\sim mv^{2}\text{.} \label{USR}%
\end{equation}
The same regions can also be considered for the photon with momentum $P-k$.
 These regions can be associated with the fields appearing in the effective
Lagrangians, see e.g. Ref.\cite{Brambilla:2004jw}. 

According to the threshold expansion prescription an integrand is expanded in
each domain to a required accuracy and the resulting integral is computed in
dimensional regularization. A detailed analysis of the full expressions in
Eqs. (\ref{Axee1}) and (\ref{Axee2}) is quite similar. To be definite  let us 
consider the integral which enters in Eq.(\ref{Axee1})
\begin{equation}
J=\int dk~\frac{e^{2}~\bar{u}_{n}\gamma_{\alpha}(l_{1}-k)\gamma_{\beta}
v_{\bar{n}}}{\left[  \left(  k-l_{1}\right)  ^{2}\right]  \left[
k^{2}\right]  \left[  \left(  k-P\right)  ^{2}\right]  }\frac{\epsilon_{\chi}^{\nu}}{4}
\text{Tr}\left[  \mathcal{P}_{1\mu\nu}\Gamma^{\alpha\beta\mu}(k)\right]  .
\label{J:def}%
\end{equation}
Keeping  the denominators of the heavy quark propagators in $\Gamma
^{\alpha\beta\mu}$ unexpanded
\begin{equation}
(p_{1}-k)^{2}-m^{2}=\left(  \frac{1}{2}P+\Delta-k\right)  ^{2}-m^{2}=k^{2}
-P_{0}k_{0}+2(\vec{k}\cdot\vec{\Delta})-\vec{\Delta}^{2}+\frac{1}{4}P_{0}
^{2}-m^{2},
\end{equation}%
\begin{equation}
(p_{2}-k)^{2}-m^{2}=\left(  \frac{1}{2}P-\Delta-k\right)  ^{2}-m^{2}=k^{2}
-P_{0}k_{0}-2(\vec{k}\cdot\vec{\Delta})-\vec{\Delta}^{2}+\frac{1}{4}P_{0}
^{2}-m^{2},
\end{equation}
where $P_{0}\sim m,$ $P_{0}^{2}/4-m^{2}\sim\left(  mv\right)  ^{2},~\ 
\vec{\Delta}\sim mv$. \ In the hard region, the small scalar products with
$\vec{\Delta}$ and the term $P_{0}^{2}/4-m^{2}\ll m^{2}~$ can be neglected resulting in 
\begin{equation}
\left[  \left(  \frac{1}{2}P\pm\Delta-k\right)  ^{2}-m^{2}\right]  _{h}\simeq
k^{2}-\left(  kP\right)  ,
\end{equation}
which appear in the  expressions (\ref{DQ}) and (\ref{DQpr})  (up to small
regularization term $\vec{\Delta}^{2}$ ).  From  dimensional counting one
immediately finds
\begin{equation}
J_{h}\sim\frac{\bar{u}_{n}\Gamma v_{n}}{m^{3}},
\label{Jh}
\end{equation}
where $\Gamma$ denotes the Dirac structure.  Computing the hard integral  $J_{h}$ in dimensional regularization one finds the IR poles $1/\varepsilon$. 
These singularities must cancel in the sum with other contribution.

Expanding the integrand (\ref{J:def}) in the soft region (\ref{SR}) yields
\begin{equation}
J_{s}\simeq\int dk~\frac{e^{2}~\bar{u}_{n}\gamma_{\alpha}\Dsl  l_{1}\gamma_{\beta
}v_{\bar{n}}}{\left[  -2\left(  kl_{1}\right)  \right]  \left[  k^{2}\right]
\left[  4m^{2}\right]  }\frac{\epsilon_{\chi}^{\nu}}{4}\text{Tr}\left[  \mathcal{P}_{1\mu\nu
}\Gamma_{s}^{\alpha\beta\mu}(k)\right]  ,
\end{equation}
where $\Gamma_{s}^{\alpha\beta\mu}(k)$ is given by (\ref{G(k)}) with
\begin{equation}
\left[  D_{Q}\right]  _{s}\sim\frac{1}{\left[  -2(k\omega)\right]  }\left\{
\gamma^{\beta}(\Dsl \omega-1)\gamma^{\alpha}+\gamma^{\alpha}(1-\Dsl \omega)\gamma
^{\beta}\right\}  ,
\end{equation}
\begin{equation}
\left[  D_{Q}^{\prime\mu}\right]  _{s}\sim\frac{1}{2m}\frac{1}{\left[
-(k\omega)\right]  }\left\{  \gamma^{\beta}\gamma^{\mu}\gamma^{\alpha}
+\gamma^{\alpha}\gamma^{\mu}\gamma^{\beta}\right\}  +\frac{1}{2m}\frac{k^{\mu
}}{\left[  -(k\omega)\right]  ^{2}}\left\{  \gamma^{\beta}(\Dsl \omega
+1)\gamma^{\alpha}-\gamma^{\alpha}(1-\Dsl \omega)\gamma^{\beta}\right\}  .\nonumber
\end{equation}
Calculating the trace and performing the contractions in the numerator results in 
\begin{equation}
J_{s}\sim\frac{1}{m^{3}}\bar{u}_{n}\Gamma_1v_{\bar{n}
}\int dk~\frac{1~}{\left[  k^{2}\right]  \left[  -(k\omega)\right]  ^{2}}
\sim\frac{\bar{u}_{n}\Gamma_{1} v_{\bar{n}}}{m^{3}}.
\label{Js}
\end{equation}
As the integral in (\ref{Js}) is scaleless  it therefore  vanishes in the dimensional regularization, i.e.  $J_{s}=0$.

In the potential region (\ref{PR}),  the expansion of the heavy quark propagator reads
\begin{equation}
\left[  \left(  \frac{1}{2}P\pm\Delta-k\right)  ^{2}-m^{2}\right]  _{p}\simeq
P_{0}^{2}/4-m^{2}-P_{0}k_{0}-\left(  \vec{k}\pm\vec{\Delta}\right)  ^{2}.
\end{equation}
The computation of the corresponding integral  then yields
\begin{align}
J_{p}  &  \simeq\frac{1}{m}\bar{u}_{n}\Gamma_1v_{\bar{n}}\int
dk~\frac{~1}{\left[  -\vec{k}^{2}\right]  \left[  P_{0}^{2}/4-m^{2}-P_{0}
k_{0}-\left(  \vec{k}+\vec{\Delta}\right)  ^{2}\right]  ^{2}}+\left(
\vec{\Delta}\rightarrow-\vec{\Delta}\right)
 \sim\bar{u}_{n}\Gamma_1v_{\bar{n}}\frac{v^{-1}}{m^{3}
}. \label{Jp}%
\end{align}
However the poles in $k_{0}$ in the integrand of Eq.(\ref{Jp})  lie in the same
imaginary half-plane   and therefore the integral
over $k_{0}$ vanishes. This observation is also true for the higher order
 contributions in $v$ appearing from this domain.  We can therefore conclude that
the potential region cannot contribute in this case.

In the ultasoft domain (\ref{USR}),  the heavy quark propagators are expanded as%
\begin{equation}
\left[  \left(  k-\frac{1}{2}P\pm\Delta\right)  ^{2}-m^{2}\right]  _{us}\simeq
P_{0}^{2}/4-m^{2}-P_{0}k_{0}-\vec{\Delta}^{2}.
\end{equation}
Performing the expansion of the integrand one gets
\begin{equation}
J_{us}\sim\frac{1}{m}\bar{u}_{n}\Gamma_1 v_{\bar{n}}\int dk~\frac
{~1}{\left[  k^{2}\right]  \left[  P_{0}^{2}/4-m^{2}-P_{0}k_{0}-\vec{\Delta
}^{2}\right]  ^{2}}\sim\bar{u}_{n}\Gamma_1 v_{\bar{n} }\frac{1}{m^{3}}.
\label{Jus}
\end{equation}
This integral  has the same scaling behavior $\sim m^{-3}$ as  the hard
integral $J_{h}$ in Eq.(\ref{Jh}).  One can also see that the integral in Eq.(\ref{Jus}) is UV divergent.
The similar analysis can also be  carried out for the second photon with momentum $k-P$. 
Therefore we conclude  that the exact integral must be given by sum
\begin{equation}
J=J_{h}+J_{us},
\end{equation}
where $J_{us}$ denotes the contributions from the both ultrasoft domains.  This
conclusion can be checked by  explicit calculations.  A similar conclusion  for the two-photon diagrams in
Fig.\ref{diagrams}  has also been obtained in Ref.\cite{Yang:2012gk}.

Guided by this consideration we suggest  that the additional relevant domain
is described by the ultrasoft region. 
In order to find the description of the
appropriate operator in the effective theory  one has to integrate out  hard
and soft photons  and leptons. After that the description of QED sector
includes only collinear leptons and ultrasoft photons. The integration of the
soft photon with the lepton and quark  must be described in the framework of
the effective theory. 

Within the above picture the factorization of the  decay amplitudes can be
described as a sum of two contributions
\begin{align}
\mathcal{A}_{J}  &  =~\bar{u}_{n}\Gamma_{J}v_{\bar{n}}~C_{\gamma\gamma}%
^{(J)}~i\left\langle \mathcal{O}(^{3}P_{0})\right\rangle \nonumber\\
&  ~\ \ \ \ \ \ \ \ +C_{\gamma}\left\langle e^{+}e^{-}\right\vert \bar{\xi
}_{n}(0)Y_{n}^{\dag}(0)\gamma_{\bot}^{\sigma}Y_{\bar{n}}(0)\xi_{\bar{n}%
}(0)~\mathcal{O}^{\sigma}(^{3}S_{1})~\left\vert \chi_{cJ}\right\rangle .
\label{A=1+2}%
\end{align}
The first term on \textit{rhs} of this equation corresponds to the hard domain
with the hard photons, $C_{\gamma\gamma}^{(J)}$ denotes the corresponding hard
coefficient function.

\begin{figure}[ptb]
\centering
\includegraphics[
height=0.7572in,
width=2.675in
]%
{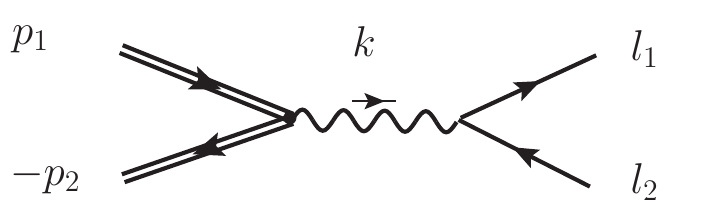}%
\caption{The hard one-photon exchange diagram.}%
\label{1photon}%
\end{figure}

The second term on \textit{rhs} \textit{ }of Eq.(\ref{A=1+2}) corresponds to
the domain with the ultrasoft photon. The operator $\mathcal{O}^{\sigma}(^{3}%
S_{1})$ is defined in Eq.(\ref{O3S1}). The outgoing collinear leptons are
described by  fields $\bar{\xi}_{n}$ and  $\xi_{\bar{n}}$ which defined as
\begin{equation}
\bar{\xi}_{n}(x)=\bar{\psi}_{c}(x)\frac{\nbs \ns}{4},~\ ~\xi_{\bar{n}}%
=\frac{\nbs \ns}{4}\psi_{c}(x).
\end{equation}
The  photon Wilson lines $Y_{n}^{\dag}$ and $Y_{\bar{n}}$ describe the
interaction of the ultrasoft longitudinal photons with the energetic lepton and
antilepton and read%
\begin{equation}
Y_{n}^{\dag}(0)=\text{Pexp}\left\{  ie\int_{0}^{\infty}ds~n\cdot
B^{us}(sn)\right\}  ,~\ \ Y_{\bar{n}}(0)=\text{\={P}exp}\left\{  -ie\int%
_{0}^{\infty}ds~\bar{n}\cdot B^{us}(s\bar{n})\right\}  ,\label{Y:def}%
\end{equation}
where $B_{\mu}^{us}$ denotes the ultrasoft photon field.  The appearance of these Wilson
lines is related with the fact that in a general gauge the tree level diagram
with attachments of $n\cdot B^{us}$ photon to the collinear field $\bar{\xi}_{n}$
describing the outgoing lepton\footnote{We assume electrical charge is
measured in proton units (positron is particle and electron is antiparticle)
that allows to use the same notation for the covariant derivative and Wilson
lines as in QCD.}  are resummed to the P-ordered exponents
\begin{equation}
\bar{\psi}_{c}\left(  1-e\Dsl{B}^{us}\frac{1}{i\Dsl{D}}\right)  \simeq\bar
{\xi}_{n}\left(  1+e~n\cdot B^{us}\frac{1}{i\left(  n\cdot D\right)  }\right)
\simeq\bar{\xi}_{n}Y_{n}^{\dag}.
\end{equation}
The leading-order hard coefficient function $C_{\gamma}$ is defined by the
diagram in Fig.\ref{1photon} and reads 
\begin{equation}
C_{\gamma}=\frac{\alpha \pi}{m^{2}}e_{Q}.\label{Cg}%
\end{equation}
The soft and collinear modes in the effective action describing the QED sector
are decoupled. This property is well known in the soft-collinear effective
theory, see e.g. Refs.\cite{Bauer:2002nz,Hill:2002vw,Beneke:2003pa}.  This  allows us to  contract the lepton fields in
the second matrix element in Eq.(\ref{A=1+2})  and rewrite it as
\begin{equation}
\left\langle e^{+}e^{-}\right\vert \bar{\xi}_{n}(0)Y_{n}^{\dag}(0)\gamma
_{\bot}^{\sigma}Y_{\bar{n}}(0)\xi_{\bar{n}}(0)~\mathcal{O}^{\sigma}(^{3}%
S_{1})~\left\vert \chi_{cJ}\right\rangle =\bar{u}_{n}\gamma_{\bot}^{\sigma
}v_{\bar{n}}~\left\langle 0\right\vert \mathcal{O}_{\gamma}^{\sigma}(^{3}%
S_{1})\left\vert \chi_{cJ}\right\rangle ,\label{O3S1YY}%
\end{equation}
with%
\begin{equation}
\mathcal{O}_{\gamma}^{\sigma}(^{3}S_{1})\equiv Y_{n}^{\dag}(0)Y_{\bar{n}%
}(0)\mathcal{O}^{\sigma}(^{3}S_{1}).
\end{equation}

The presence of the soft scale $\vec{\Delta}^{2}$ in Eqs.~(\ref{A1}) and
(\ref{A2}) can be explained by the contribution with ultrasoft photon.  Therefore
 in order to find the hard coefficient functions $C_{\gamma\gamma}^{(J)}$ we
have to perform the matching onto the configuration  described by
Eq.(\ref{A=1+2}).  For that purpose we need to compute the ultrasoft matrix element
(\ref{O3S1YY}) in the effective theory.

\begin{figure}[ptb]
\centering
\includegraphics[
height=1.4512in,
width=3.3508in
]%
{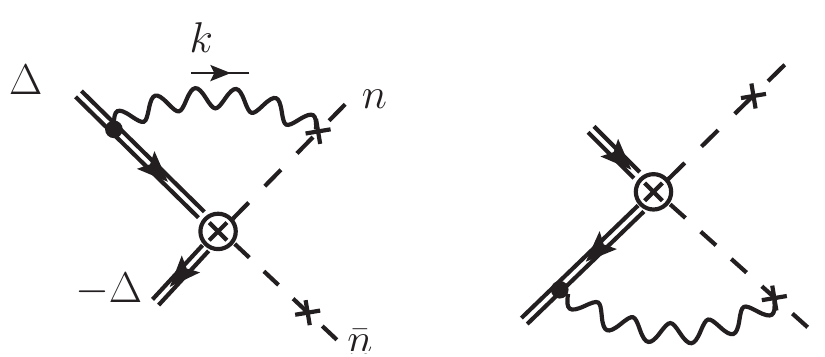}
\caption{The diagrams generated by the $T$-product (\ref{TOg3S1}) in pNRQED.  The crossed
circle denotes the vertex of the operator $O_{\gamma}(^{3}S_{1})$, the dashed
lines represent the Wilson lines associated with light-like directions $n$ and
$\bar{n}$. The small crosses on the dashed lines show all possible attachments of the photon.  }%
\label{pnrqed-me}%
\end{figure}

The interaction of ultrasoft photons with quarks are described within the pNRQED.
The ultrasoft photons have momentum $p\sim mv^{2}$ so that photon field scales as
\begin{equation}
B_{\mu}^{us}\sim mv^{2}.
\end{equation}
The scaling of the quark fields reads%
\begin{equation}
\psi_{\omega}\sim(mv)^{3/2},~\ \vec{\partial}_{i}\psi_{\omega}\sim
(mv)\psi_{\omega},~\ \partial_{0}\psi_{\omega}\sim(mv^{2})\psi_{\omega}.
\end{equation}
Using this counting one finds
\begin{equation}
C_{\gamma\gamma}^{(J)}~\mathcal{O}^{\sigma}(^{3}P_{J})\sim m^{-3}(mv)^{4}.
\label{CO3PJ}%
\end{equation}
At the same time \
\begin{equation}
C_{\gamma}\mathcal{O}_{\gamma}^{\sigma}(^{3}S_{1})\sim C_{\gamma}%
\mathcal{O}^{\sigma}(^{3}S_{1})\sim m^{-2}(mv)^{3}.
\end{equation}
However the pure quark operator $\mathcal{O}^{\sigma}(^{3}S_{1})$ is $C$-odd and
therefore it cannot contribute to the matrix element with a $C$-even charmonium state
\begin{equation}
\left\langle 0\right\vert \mathcal{O}^{\sigma}(^{3}S_{1})\left\vert \chi
_{cJ}\right\rangle =0.
\end{equation}
In order to obtain a nontrivial contribution one needs to consider at least one interaction  of an ultrasoft
photon with the quark in pNRQED.  We only need the two-particle sector describing the
electromagnetic interactions of quarks (in rest frame~$\omega=(1,0)$)
\begin{equation}
\mathcal{L}_{0}^{em}[B^{us}]=\int d^{4}x~\psi_{\omega}^{\dag}(x)\gamma_{0}\left(
i\omega\cdot\partial+\frac{i\partial_{\top}\cdot i\partial_{\top}}{2m}\right)
\psi_{\omega}(x),~~ \label{L0}%
\end{equation}%
\begin{equation}
\mathcal{L}_{1}^{em}[B^{us}]=\int d^{4}x~\psi_{\omega}^{\dag}(x)\gamma_{0}\left[
~\vec{x}\cdot\partial_{\top}~ee_{Q}~\omega\cdot B^{us}(x_{0})+\frac{1}%
{m}ee_{Q}B^{us}(x_{0})\cdot i\partial_{\top}\right]  ~\psi_{\omega}(x),
\label{L1}%
\end{equation}
and analogous contributions with antiquark fields. The arguments of the ultrasoft
photon field are expanded because the space components of the quark fields
varies at $\vec{x} \sim1/mv$, the measure scales as  $dx_{0}\sim
1/mv^{2},~\ d^{3}\vec{x}\sim (mv)^{-3}$. With these rules  one finds that
$\mathcal{L}_{0}^{em}\sim v^{0}$ and $\mathcal{L}_{1}^{em}\sim v^{1}$.  The
leading-order term (\ref{L0}) provides the soft quark propagator
\begin{equation}
\Delta_{\omega}(k)=\frac{i}{(\omega k)-\vec{k}^{2}/2m+i\varepsilon}.
\end{equation}
A nontrivial contribution to the matrix element $\left\langle 0\right\vert
...$\ $\left\vert \chi_{cJ}\right\rangle$ can be obtained from $T$-product
\begin{equation}
T\{\mathcal{O}_{\gamma}^{\sigma}(^{3}S_{1}),\mathcal{L}_{1}^{em}[B^{us}]\}\sim
m v^{4},
\label{TOg3S1}
\end{equation}
which is of the same order as the hard contribution in  Eq.(\ref{CO3PJ}). Calculation of this $T$-product
gives diagrams shown in Fig.\ref{pnrqed-me}. The dashed lines can be
associated with the collinear leptons or equivalently with the ultrasoft Wilson
lines (\ref{Y:def}).
These diagrams induce a mixing of the operators $O_{\gamma}^{\sigma}(^{3}%
S_{1})$ and $\mathcal{O}(^{3}P_{J})$ due to electromagnetic interaction in the
framework of pNRQED. 

In order to perform the matching onto  operators according to formula
(\ref{A=1+2}) one has also to compute the contribution of the diagrams in
Fig.\ref{pnrqed-me}. The simplest way to proceed is to follow the same
technique as we used above for diagrams in Fig.\ref{diagrams}.

Let us consider
$\chi_{c2}$ as initial state. In  this calculation we set $P_{0}=2m$
and only keep the relative momentum $\vec{\Delta}$.  Then the sum of all
four diagrams gives
\begin{equation}
\left\langle e^{+}e^{-}\right\vert C_{\gamma}T\{\mathcal{O}_{\gamma}^{\sigma
}(^{3}S_{1}),\mathcal{L}_{1}^{em}[B^{us}]\}\left\vert \chi_{c2}\right\rangle
=\bar{u}_{n}\Gamma_{2}v_{\bar{n}}~i\left\langle \mathcal{O}(^{3}%
P_{0})\right\rangle ~8C_{\gamma}\frac{e^{2}e_{Q}}{2m}~J_{us}.
\end{equation}
Computing these diagrams we project the soft quarks fields on the operator
$\mathcal{O}(^{3}P_{2})$ and substitute the corresponding matrix element which
gives the factor $~\epsilon_{\chi}^{\sigma\rho}i\left\langle \mathcal{O}(^{3} P_{0})\right\rangle $, 
the coefficient $4$ arises from the sum of the four
diagrams shown in Fig.\ref{pnrqed-me}, the ultrasoft loop integral reads
\begin{equation}
~J_{us}=(-i)\int dk~\frac{1}{\left[  k^{2}\right]  }\frac{1}{\left[  -(\omega
k)-\vec{\Delta}^{2}/2m\right]  ^{2}}. \label{Jus:PNRQCD}%
\end{equation}
This integral coincides with the ultrasoft integral  of Eq.(\ref{Jus}) obtained within the
threshold expansion approach up to term $P^{2}_{0}/4-m^{2}$ which vanishes because
we set $P_{0}=2m$. The  integral in Eq.(\ref{Jus:PNRQCD}) is UV-divergent and
we use dimension regularization $D=4-2\varepsilon$ in order to compute it.
 The result reads
\begin{equation}
J_{us}=\frac{\pi^{D/2}}{(2\pi)^{D}}~\left(  -\frac{2}{\varepsilon}\right)
\left(  \frac{\vec{\Delta}^{2}}{m\mu_{F}}\right)  ^{-2\varepsilon},
\label{JUS}%
\end{equation}
where $\mu_{F}$ is the factorization scale. 
The $1/\varepsilon$ pole is the UV-pole  which describes UV-mixing of the
operators $O_{\gamma}^{\sigma}(^{3}S_{1})$ and $\mathcal{O}^{\sigma}(^{3}%
P_{J})$, schematically%
\begin{equation}
\left[  O_{\gamma}(^{3}S_{1})\right]  _{R}=O_{\gamma}(^{3}S_{1})+Z_{J}%
~\mathcal{O}(^{3}P_{J}), \label{OR}%
\end{equation}
where $\left[  \mathcal{O}\right]  _{R}$ on the \textit{lhs} of Eq.(\ref{OR}) denotes
the renormalized operator. Furthermore,  $Z_{J}\sim e^{2}/\varepsilon$ is the corresponding
renormalization constant.   Assuming $\overline{MS}$-subtraction scheme one
finds
\begin{equation}
\left[  J_{us}\right]  _{R}=\frac{1}{4\pi^{2}}\ln\frac{\vec{\Delta}^{2}}%
{m\mu_{F}}.
\end{equation}
Hence we obtain
\begin{equation}
\left\langle e^{+}e^{-}\right\vert C_{\gamma}T\{\mathcal{O}_{\gamma}^{\sigma
}(^{3}S_{1}),\mathcal{L}_{1}^{em}[B^{us}]\}\left\vert \chi_{c2}\right\rangle
_{R}=\bar{u}_{n}\Gamma_{2}v_{\bar{n}}~i\left\langle \mathcal{O}(^{3}%
P_{0})\right\rangle ~\frac{\alpha^{2}}{m^{3}}e_{Q}^{2}~4\ln\frac{\vec{\Delta
}^{2}}{m\mu_{F}}.
 \label{O3gmeXc2}%
\end{equation}
The soft matrix element for the $\chi_{c1}$ can be computed in the same way,
resulting  in 
\begin{equation}
\left\langle e^{+}e^{-}\right\vert C_{\gamma}T\{\mathcal{O}_{\gamma}^{\sigma
}(^{3}S_{1}),\mathcal{L}_{1}^{em}[B^{us}]\}\left\vert \chi_{c1}\right\rangle
_{R}=\bar{u}_{n}\Gamma_{1}v_{\bar{n}}~i\left\langle \mathcal{O}(^{3}%
P_{0})\right\rangle ~\frac{\alpha^{2}}{m^{3}}e_{Q}^{2}~2\sqrt{2}\ln\frac
{m\mu_{F}}{\vec{\Delta}^{2}}. \label{O3gmeXc1}%
\end{equation}

The hard coefficients $C_{\gamma\gamma}^{(J)}$ are given by 
\begin{equation}
C_{\gamma\gamma}^{(J)}=\frac{\mathcal{A}_{J}-C_{\gamma}\bar{u}_{n}\gamma
_{\bot}^{\sigma}v_{\bar{n}}\left\langle 0\right\vert \mathcal{O}_{\gamma
}^{\sigma}(^{3}S_{1})\left\vert \chi_{cJ}\right\rangle }{\bar{u}_{n}\Gamma
_{J}v_{\bar{n}}~\left\langle 0\right\vert \mathcal{O}^{\sigma}(^{3}%
P_{J})\left\vert \chi_{cJ}\right\rangle }, \label{CggJ}%
\end{equation}
where the expressions for $\mathcal{A}_{J}$ are given by Eqs.(\ref{A1}) and (\ref{A2}).  
The  important check of  the factorization formula (\ref{A=1+2}) is the
cancellation of the ultrasoft scale $\vec{\Delta}^{2}$ in the expressions for $C_{\gamma\gamma}^{(J)}$ obtained from Eq.(\ref{CggJ}).
Substituting the computed expressions in Eq.(\ref{CggJ}) we obtain
\begin{equation}
~C_{\gamma\gamma}^{(1)}=\frac{\alpha^{2}}{m^{3}}e_{Q}^{2}~\sqrt{2}\ln
\frac{m^{2}}{4\mu_{F}^{2}},
\label{C1gg}
\end{equation}%
\begin{equation}
C_{\gamma\gamma}^{(2)}=~\frac{\alpha^{2}}{m^{3}}e_{Q}^{2}2\left\{  \ln\frac
{\mu_{F}^{2}}{m^{2}}+\frac{2}{3}\left(  \ln2-1+i\pi\right)  \right\}  .
\label{C2gg}
\end{equation}
These expressions are the main result of this section. We observe that the soft scale cancel  in
Eqs.(\ref{C1gg}) and (\ref{C2gg})  as it must be.  Hence  the
 factorization formula described by Eq.(\ref{A=1+2})  describes properly the ultrasoft region of the
one-loop diagram.

The coefficient function $C_{\gamma\gamma}^{(2)}$ has an imaginary part which
originates from the two-photon cut. Such mechanism can not work for $\chi
_{c1}$ state, therefore $C_{\gamma\gamma}^{(1)}$ is real.

The hard coefficient functions depend on the factorization scale $\mu_{F}$.
Therefore
\begin{equation}
\mathcal{A}_{J}=\bar{u}_{n}\Gamma_{J}v_{\bar{n}}~C_{\gamma\gamma}^{(J)}%
(\mu_{F})~i\left\langle \mathcal{O}(^{3}P_{0})\right\rangle
 +C_{\gamma}~\bar{u}_{n}\gamma^{\sigma}v_{\bar{n}%
}\left\langle 0\right\vert \mathcal{O}_{\gamma}^{\sigma}(^{3}S_{1}%
)\left\vert \chi_{cJ}\right\rangle (\mu_{F}), \label{AJmuF}%
\end{equation}
and the independence of the amplitude $\mathcal{A}_{J}$  on $\mu_{F}$ yields the evolution equation
\begin{equation}
\bar{u}_{n}\Gamma_{J}v_{\bar{n}}
i\left\langle \mathcal{O}(^{3}P_{0})\right\rangle ~\mu_{F}\frac{d}%
{d\mu_{F}}C_{\gamma\gamma}^{(J)}(\mu_{F})=-C_{\gamma}\bar{u}_{n}\gamma
^{\sigma}v_{\bar{n}}~~\mu_{F}\frac{d}{d\mu_{F}}\left\langle 0\right\vert
\mathcal{O}_{\gamma}^{\sigma}(^{3}S_{1})\left\vert \chi_{cJ}\right\rangle
(\mu_{F}).
\end{equation}
The solution of this equation depends on the initial condition defined at some scale $\mu_{0}$. 
 Performing  numerical estimates one has to fix a value of this scale. By
derivation this scale separates the  hard region (two hard photons) from the
ultrasoft region (hard and ultrasoft photons).  Therefore it is natural to
 associate this scale with the virtuality of the ultrasoft photon and to set
$\mu_{0}$ to be of order $300-500$ MeV.  
Then the 
matrix element of the operator $\mathcal{O}_{\gamma}^{\sigma}(^{3}S_{1})$  on   the  \textit{rhs} of Eq.(\ref{AJmuF}) describes only the ultrasoft
 nonperturbative contribution which can be only estimated  within some
low-energy effective theory or model.  Similar to the 
well known   color octet mechanism,  the operator $\mathcal{O}_{\gamma}^{\sigma}(^{3}S_{1})$ can also be associated with
the   electromagnetic  mechanism.  
The corresponding  matrix element can be interpreted as an  overlap with the higher Fock state  $|Q\bar Q \gamma \rangle$ which includes 
a dynamical photon  while the matrix elements  of the operators $\mathcal{O}(^{3}P_{J})$ describe the coupling  to the dominant  quark-antiquark state. 
 Therefore the full description of the leptonic decay  requires  a knowledge on the subleading structure of the quarkonium state.  

\begin{figure}[ptb]%
\centering
\includegraphics[
height=1.3026in,
width=3.0544in
]%
{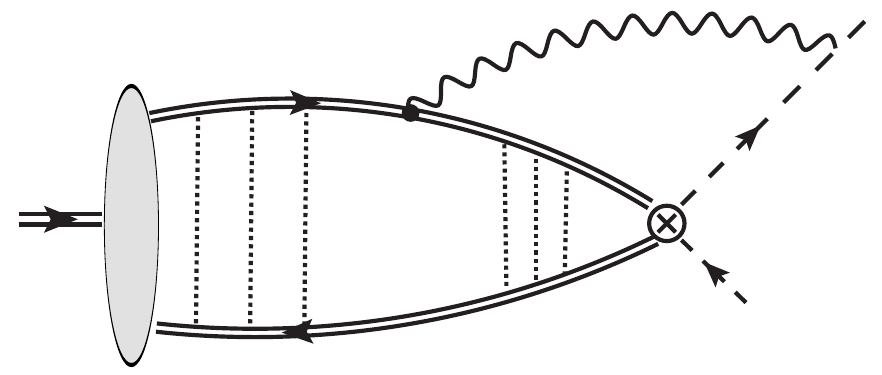}%
\caption{An example of the diagram in pNRQCD describing the ultrasoft matrix element
of Eq. (\ref{O3S1YY}).  The doted lines denote potential gluons with momenta
given by Eq.(\ref{PR}). }%
\label{pnrqed-coulomb}%
\end{figure}

In the large mass limit $m\rightarrow\infty$ one can consider a specific
situation known as the Coulomb limit when the binding energy 
 is larger then the typical hadronic scale $E\sim$ $mv^{2}\gg\Lambda_{\text{QCD}}$. In this
case the strong coupling is quite small $\alpha_{s}(mv)\sim v$ and ultrasoft
contribution can be estimated within the pNRQCD. Then one has to
compute the diagram as in Fig.\ref{pnrqed-coulomb}  resumming the
interactions with Coulomb gluons.
Such calculation has been carried out for the radiation function in
Ref.\cite{Beneke:1999gq}. Perhaps, such calculation might also be interesting
here in order to get an idea about the relative value of this matrix element in the  Coulomb limit. In
present paper we will obtain an estimate of the ultrasoft
matrix element using  the so-called heavy hadron chiral perturbation theory (HH$\chi$PT) framework  in the next section.

\newpage

\section{\bigskip Phenomenology}
\label{phen}

\subsection{Calculation of the ultrasoft matrix element in the heavy hadron chiral
perturbation theory }

In order to provide a numerical estimate of the decay rate we need to estimate
the ultrasoft matrix elements
\begin{equation}
\left\langle 0\right\vert \mathcal{O}_{\gamma}^{\sigma}(^{3}S_{1})\left\vert
\chi_{cJ}\right\rangle \label{meO3S1g}%
\end{equation}
which describes an overlap  with the higher Fock component  of the
charmonium state $\chi_{cJ}$ in which a dynamical photon is present. 
One can expect that the soft photon has already
quite large wavelength and therefore it interacts with heavy charmonium as
with a point-like source. Then it is natural to expect that the relevant
dynamical degrees of freedom in this case are associated with mesonic fields
and the corresponding low energy dynamics is described by the most generic
effective action compatible with the symmetries of NRQCD. Such an approach
is known as heavy hadron chiral perturbation theory  in
Refs.\cite{Wise:1992hn, Burdman:1992gh} for the heavy-light mesons and then
generalized on quarkonia in Refs.\cite{Casalbuoni:1992yd, Mannel:1994xh,
Casalbuoni:1996pg}. This framework can also be used for the calculation of
the  matrix element in Eq.(\ref{meO3S1g}).

For our purpose we need only the electromagnetic sector of the HH$\chi$PT
described by the effective action which includes the kinetic terms for $J/\psi$ and $\psi'$
states and the vertices  describing the  electromagnetic vertices $\chi_{cJ}\, J/\psi\,\gamma$ and $\chi_{cJ}\, \psi'\, \gamma$ 
\footnote{We are grateful to Maxim Polyakov for discussion of the contribution with the virtual state $\psi'$. }.  
As before we assume the rest frame for the initial state
$\chi_{cJ}$. The kinetic  Lagrangian reads
\begin{equation}
\mathcal{L}_{\text{kin}}(x)=\frac{1}{2}2M_{\chi}~\psi_{\mu}^{(\omega
)}(x)\left\{  i(\omega\partial)-\Delta M\right\}  \psi_{\mu}^{(\omega)}(x)+
\frac{1}{2}2M_{\chi}~\psi_{\mu}^{\prime( \omega
)}(x)\left\{  i(\omega\partial)-\Delta' M\right\}  \psi_{\mu}^{\prime(\omega)}(x),
\label{Lkin}%
\end{equation}
with the residual masses $\Delta M=(M^{2}_{\chi}-M^{2}_{\psi})/2M_{\chi}$ and $\Delta' M=(M^{2}_{\chi}-M^{2}_{\psi'})/2M_{\chi}$. 
The fields $\psi_{\mu}^{(\omega)}$ and  $\psi_{\mu}^{\prime (\omega)}$ describes the residual motion of the heavy $J/\psi$ and $\psi'$ particles
and satisfy $\omega^{\mu}\psi_{\mu}^{(\omega)}(x)=\omega^{\mu}\psi_{\mu}^{\prime (\omega)}(x)=0$. 

The leading-order in $1/m$ effective Lagrangian describing the radiative decays
$\chi_{cJ}\rightarrow J/\psi+\gamma$ and $\psi'\rightarrow \chi_{cJ}+\gamma$ reads \cite{Casalbuoni:1996pg}
\begin{equation}
\mathcal{L}^{em}_{SP}=\frac{1}{2}ee_{Q}f_{\gamma}~\text{Tr}\left[  \gamma
_{0}\text{~}J_{S}^{\dag}\gamma_{0}J_{P}^{\mu}\right]  F_{\mu\nu}\omega^{\nu}+\mathcal{L}^{em}_{SP}+\frac{1}{2}ee_{Q}f'_{\gamma}~\text{Tr}\left[  \gamma
_{0}\text{~}J_{S}^{\prime \dag}\gamma_{0}J_{P}^{\mu}\right]  F_{\mu\nu}\omega^{\nu}+h.c.~
\label{Lem}%
\end{equation}
with%
\bea
J_{S}=\frac{1}{2}(1+\Dsl\omega)\left\{  \psi_{\alpha}^{(\omega)}\gamma
^{\alpha}-\eta_{c}\gamma_{5}\right\}  \frac{1}{2}(1-\Dsl\omega),
\\
J'_{S}=\frac{1}{2}(1+\Dsl\omega)\left\{  \psi_{\alpha}^{\prime(\omega)}\gamma
^{\alpha}-\eta'_{c}\gamma_{5}\right\}  \frac{1}{2}(1-\Dsl\omega),
\eea
and
\begin{equation}
J_{P}^{\mu}=\frac{1}{2}(1+\Dsl\omega)\left\{ - \chi_{2}^{\mu\alpha}\gamma_{\alpha
}-\frac{1}{\sqrt{2}}i\varepsilon^{\mu\alpha\beta\rho}\gamma_{\alpha}\chi
_{1\beta}\omega_{\rho}+\frac{1}{\sqrt{3}}\left(  \gamma^{\mu}-\omega^{\mu
}\right)  \chi_{0}+h_{c}^{\mu}\gamma_{5}\right\}  \frac{1}{2}(1-\Dsl\omega).
\end{equation}
The currents $J_{S}, J'_{S}$ and $J_{P}^{\mu}$ describe  particles from $S$- and $P$-wave multiplets, respectively. 
In Eq.(\ref{Lem}) we introduced the
dimensionless couplings  $f_{\gamma}$ and $f'_{\gamma}$. The fields $\chi_{J}$ 
describe charmonium states $\chi_{cJ}$.  Computing
the trace in Eq.(\ref{Lem})  one finds
\bea
\mathcal{L}^{em}_{SP}=-ee_{Q}f_{\gamma}~\chi_{2}^{\mu\alpha}\psi_{\alpha}^{(\omega)}F_{\mu\nu
}\omega^{\nu}-\frac{ee_{Q}f_{\gamma}~}{\sqrt{2}}i\varepsilon^{\mu\alpha
\beta\rho}\psi^{(\omega)}_{\alpha}\chi_{1\beta}\omega_{\rho}F_{\mu\nu}\omega^{\nu
}
\\
-ee_{Q}~f'_{\gamma}~\chi_{2}^{\mu\alpha}\psi_{\alpha}^{\prime(\omega)}F_{\mu\nu
}\omega^{\nu}-\frac{ee_{Q}f'_{\gamma}~}{\sqrt{2}}i\varepsilon^{\mu\alpha
\beta\rho}\psi^{\prime(\omega)}_{\alpha}\chi_{1\beta}\omega_{\rho}F_{\mu\nu}\omega^{\nu}
+~...
\eea
where we show only the relevant terms. 

\begin{figure}[ptb]%
\centering
\includegraphics[
height=0.9431in,
width=4.6633in
]%
{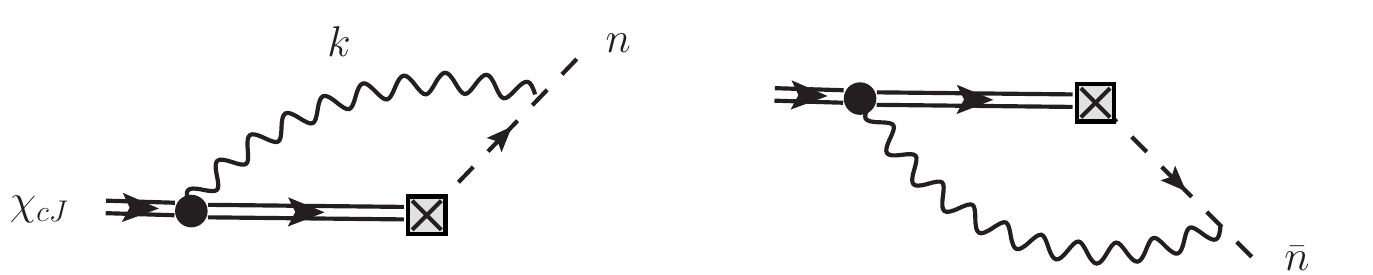}%
\caption{The diagrams which describe the matrix element (\ref{Og3S1ChPT}) in
HH$\chi$PT. The crossed box denotes the operator (\ref{psiYY}), dashes lines
describe the Wilson lines $Y_{n}^{\dag}$ and $Y_{\bar{n}}$, black circle
corresponds to the interaction vertices generated by $\mathcal{L}^{em}_{SP}$
(\ref{Lem}).  }%
\label{hhchpt-graph}%
\end{figure}

Our calculations involve operators $\mathcal{O}(^{3}P_{J})$ and $\mathcal{O}%
(^{3}S_{1})$ which have also to be matched onto physical quarkonium fields.
The spin symmetry in the heavy quark limit yields
\bea
\left[  \mathcal{O}^{\sigma}(^{2s+1}S_{1})\right]  _{\alpha\beta}=\left\langle
\mathcal{O}(^{3}S_{1})\right\rangle \left[  J\right]  _{\alpha\beta}
+\left\langle\mathcal{O}'(^{3}S_{1})\right\rangle \left[  J'\right]  _{\alpha\beta},
\label{Jab}%
\eea
\begin{equation}
\left[  \mathcal{O}^{\mu}(^{2s+1}P_{J})\right]  _{\alpha\beta}=\left\langle
\mathcal{O}(^{3}P_{0})\right\rangle \left[  J^{\mu}\right]  _{\alpha\beta
},\label{Jmuab}%
\end{equation}
where $\alpha\beta$ are spinor indices. Taking the matrix element  and
computing the traces one can see that Eqs.(\ref{Jab}) and (\ref{Jmuab})
reproduce correctly the matrix elements (\ref{me3S1})-(\ref{me3P2}).  Using
these results one finds 
\begin{equation}
\mathcal{O}^{\sigma}_{\gamma}(^{3}S_{1})
 \simeq \left\{ 
\left\langle \mathcal{O}(^{3}S_{1})\right\rangle \psi^{(\omega)\sigma}(0)
 +\left\langle \mathcal{O}'(^{3}S_{1})\right\rangle \psi^{\prime(\omega)\sigma}(0)
\right\}
Y_{n}^{\dag}Y_{\bar{n}}.
\label{psiYY}%
\end{equation}
Hence%
\begin{equation}
\left\langle 0\right\vert ~\mathcal{O}_{\gamma}^{\sigma}(^{3}S_{1})~\left\vert
\chi_{cJ}\right\rangle =
\left\langle 0\right\vert ~T\left\{ \left( \left\langle \mathcal{O}(^{3}S_{1})\right\rangle \psi^{(\omega)\sigma}(0)
+\left\langle \mathcal{O}'(^{3}S_{1})\right\rangle \psi^{\prime(\omega)\sigma}(0) \right )Y_{n}^{\dag}Y_{\bar{n}%
},\mathcal{L}^{em}_{SP}\right\}  ~\left\vert \chi_{cJ}\right\rangle
.\label{Og3S1ChPT}%
\end{equation}
Computing the  $T$-product in Eq.(\ref{Og3S1ChPT}) gives the diagrams in
Fig.\ref{hhchpt-graph}.
These diagrams are UV-divergent and we use  in our calculation the dimensional regularization and
$\overline{MS}$ subtraction scheme. The results read
\begin{equation}
\left\langle 0\right\vert ~\mathcal{O}_{\gamma}^{\sigma}(^{3}S_{1})~\left\vert
\chi_{c1}\right\rangle =i\varepsilon_{\perp}^{\sigma\alpha}\epsilon_{\chi
\alpha}~i\frac{\alpha}{\pi}e_{Q}~\frac{1}{\sqrt{2}}h(\mu_{\chi}),
\label{meOg1ChPT}%
\end{equation}%
\begin{equation}
\left\langle 0\right\vert ~\mathcal{O}_{\gamma}^{\sigma}(^{3}S_{1})~\left\vert
\chi_{c2}\right\rangle =\epsilon_{\chi}^{\alpha\sigma}n_{\alpha}~i\frac
{\alpha}{\pi}e_{Q}~h(\mu_{\chi}),
\label{meOg2ChPT}%
\end{equation}
where $i\varepsilon_{\perp}^{\sigma\alpha}\equiv i\varepsilon^{\sigma
\alpha\beta\rho}n_{\beta}\bar{n}_{\rho}/2$ and
\begin{equation}
h(\mu_{\chi})=f_{\gamma}\left\langle \mathcal{O}(^{3}S_{1})\right\rangle
\frac{\Delta M}{M_{\chi}}\left(  \ln2-1-\ln\frac{\mu_{\chi}}{\Delta M}-i\pi \right)+ 
f'_{\gamma}\left\langle \mathcal{O}'(^{3}S_{1})\right\rangle
\frac{\Delta' M}{M_{\chi}}\left(  \ln2-1-\ln\frac{\mu_{\chi}}{-\Delta' M} \right) .
\label{hChPT}%
\end{equation}
From these results  one sees that the spin symmetry of NRQCD relates the soft
photon matrix elements for $J=1$ and $J=2$ which are defined by the same nonperturbative
couplings $f_{\gamma}\left\langle \mathcal{O}(^{3}S_{1})\right\rangle$ and $f'_{\gamma}\left\langle \mathcal{O}'(^{3}S_{1})\right\rangle$. 
The  imaginary part in Eq.(\ref{hChPT}) corresponds to the photon-quarkonium ($J/\psi$) cut in the
diagrams in Fig.\ref{hhchpt-graph}.  Hence we conclude  that the two-photon cut
appears only in the hard photon contribution.  The contribution with $\psi'$ has no physical cut and in the diagram this is provided by the negative value of $\Delta' M$.

The UV-poles in the HH$\chi$PT diagrams appear due to the mixing of the
operators $\mathcal{O}_{\gamma}^{\sigma}(^{3}S_{1})$ and $\chi_{cJ}$.
Therefore this UV-pole can be absorbed into renormalization of the chiral
constant in front of the operators $\chi_{cJ}$\footnote{Or equivalently one can say that this pole renormalizes the contact vertex describing the $\chi_{cJ}\rightarrow e^{+}e^{-}$ decay}.  The expression for the total
amplitude now reads
\begin{equation}
\mathcal{A}_{1}=~i\bar{u}_{n}\Gamma_{1}v_{\bar{n}}~\left\{  ~\tilde{C}_{1}%
(\mu_{\chi})-C_{\gamma}\frac{\alpha}{\pi}e_{Q}\frac{1}{\sqrt{2}}h(\mu_{\chi
})\right\}  , \label{A1 HHChPT}%
\end{equation}
\begin{equation}
\mathcal{A}_{2}=~i\bar{u}_{n}\Gamma_{2}v_{\bar{n}}~\left\{  \tilde{C}_{2}%
(\mu_{\chi})+C_{\gamma}\frac{\alpha}{\pi}e_{Q}~h(\mu_{\chi})~~\right\}  .
\label{A2 HHChPT}%
\end{equation}
We set  the value of the chiral scale $\mu_{\chi}=\mu_{0}$,  defining the
chiral couplings $\tilde{C}_{J}(\mu_{0})$ as product of the two-photon hard
coefficient functions $C_{\gamma\gamma}^{(J)}$ and constant $\left\langle
\mathcal{O}(^{3}P_{0})\right\rangle $ 
\begin{equation}
\tilde{C}_{J}(\mu_{0})=C_{\gamma\gamma}^{(J)}(\mu_{0})~\left\langle \mathcal{O}(^{3}P_{0})\right\rangle .
\end{equation}
This defines the expressions for the amplitudes which will be  used for
our numerical estimates.

\subsection{\bigskip Numerical estimates}

In order to perform  numerical estimates we need the values of the
nonperturbative parameters $\left\langle \mathcal{O}(^{3}S_{1})\right\rangle$,
 $\left\langle \mathcal{O}(^{3}P_{0})\right\rangle $ and $f_{\gamma}$.
Two of them are related to the values of quarkonium wave function at the
origin, see Eqs.(\ref{def:R10}) and  (\ref{def:R21}). Their absolute values have
been estimated in Ref. \cite{Eichten:1995ch} using  different models for the potential. In our numerical calculations we use 
 the values obtained for Buchm\"uller-Tye potential \cite{Buchmuller:1980su}
\begin{equation}
|R_{21}^{\prime}(0)|^{2}\simeq0.075\text{GeV}^{5},
\end{equation}%
\begin{equation}
\left\vert R_{10}(0)\right\vert ^{2}\simeq0.81\text{GeV}^{3},  \quad \left\vert R_{20}(0)\right\vert ^{2}\simeq0.530\text{GeV}^{3}
\end{equation}
We also assume that they correspond to positive values:
\begin{equation}
R_{10}(0)>0,\quad R_{20}(0)>0, \quad R_{21}^{\prime}(0)>0.  
\label{signR}
\end{equation}

The absolute values of the electromagnetic couplings $f_{\gamma}$ and  $f'_{\gamma}$ can be
estimated from the decays $\chi_{cJ}\rightarrow J/\psi\gamma$ and $\psi'\rightarrow \chi_{cJ}\gamma$.  Using for
widths $\Gamma\lbrack\chi_{c1}]=0.84\times10^{-3}$GeV,   $\Gamma\lbrack
\chi_{c2}]=1.93\times10^{-3}$GeV and branching fractions $Br[\chi
_{c1}\rightarrow J/\psi\gamma]=0.340$ , $Br[\chi_{c2}\rightarrow J/\psi
\gamma]=0.192$   from \cite{Agashe:2014kda}  we obtain
\begin{equation}
\left\vert f_{\gamma}\right\vert =\sqrt{\frac{\Gamma\lbrack\chi_{cJ}%
]Br[\chi_{cJ}\rightarrow J/\psi\gamma]}{\frac{1}{2}\alpha e_{Q}^{2}~k_{0}%
^{3}/M_{\chi_{cJ}}^{2}}}\simeq\left\{
\begin{array}
[c]{c}%
5.87\left(  \chi_{c0}\right)\\
6.05\left(  \chi_{c1}\right)  \\
6.03\left(  \chi_{c2}\right)
\end{array}
\right\}  \simeq 6.0,
\end{equation}
where $k_{0}=(M_{\chi_{cJ}}^{2}-M_{\psi}^{2})/2M_{\chi_{cJ}}$ is the photon
energy.  Similarly,  using 
widths $\Gamma\lbrack \psi']=0.299\times10^{-3}$GeV  and branching fractions $Br[\psi'
\rightarrow \chi_{c1}\gamma]=0.096$ , $Br[\psi'\rightarrow \chi_{c2}
\gamma]=0.091$   from \cite{Agashe:2014kda}  we obtain
\begin{equation}
\left\vert f'_{\gamma}\right\vert=\left\{
\begin{array}
[c]{c}%
6.5 \left(  \chi_{c0}\gamma\right)  \\
7.0 \left(  \chi_{c1}\gamma\right)  \\
8.1 \left(  \chi_{c2}\gamma\right)
\end{array}
\right\}  \simeq 7.2,
\end{equation}

 We also need to know the sign of this coupling
which   can only be defined by a specific nonperturbative calculation.  It
turns out  that this coupling   can be represented as an overlap
integral of the radial wave functions.  Comparing our results for the decay
amplitudes $\chi_{cJ}\rightarrow J/\psi\gamma$  with  the ones  computed in
Ref.\cite{Brambilla:2012be}   we find
\begin{equation}
f_{\gamma}=\sqrt{2M_{\chi}}\sqrt{2M_{\psi}}\frac{1}{\sqrt{3}}\int%
_{0}^{\infty}drr^{3}R_{21}(r)R_{10}(r),
\label{foverlap}
\end{equation}
where factors $\sqrt{2M}$ appear  due to relativistic normalizations of
the hadronic states.  The analogous expression also holds for the coupling $f'_{\gamma}$. The overlap integral has been computed in the
framework of potential models, see e.g. Refs.\cite{Eichten:1978tg,Brambilla:2004wf}. Its
value is found to be positive for $f_{\gamma}$ and negative for $f'_{\gamma}$.  Therefore we assume in the following  that  $f_{\gamma}>0$ and $f'_{\gamma}<0$. 

The  expressions for the decay  width read
\begin{equation}
\Gamma\lbrack\chi_{cJ}\rightarrow e^{+}e^{-}]=\left\{
\begin{array}
[c]{c}%
\frac{1}{12\pi}M_{\chi}~|C_{\gamma\gamma}^{(1)}(\mu_{0})\left\langle \mathcal{O}(^{3}P_{0})\right\rangle-C_{\gamma}%
\frac{\alpha}{\pi}e_{Q}h(\mu_{0})/\sqrt{2}~|^{2}\\
\frac{1}{40\pi}M_{\chi}~|C_{\gamma\gamma}^{(2)}(\mu_{0})\left\langle \mathcal{O}(^{3}P_{0})\right\rangle+C_{\gamma}%
\frac{\alpha}{\pi}e_{Q}h(\mu_{0})~|^{2}%
\end{array}
\right.  ,\label{GammaXee}%
\end{equation}
where $C_{\gamma}$ is given by  Eq.(\ref{Cg}).  We use  $m_{c}=1.5$~GeV for
the mass of the charm quark  and compute $h(\mu_{0})$ by substituting  $M_{\chi
}=(M_{\chi_{c1}}+M_{\chi_{c2}})/2$ in the expression (\ref{hChPT}). 

Our numerical 
results are presented  in Table~\ref{tab:results} for different  values of $\mu_{0}$.  
\begin{table}[ht]
\caption{Numerical results for the decay widths for different values of the factorization scale $\mu_{0}$.}
\centering%
\begin{tabular}
[c]{|c|c|c|}\hline
$\mu_{0},\text{MeV}$ & $\Gamma\lbrack\chi_{c1}\rightarrow e^{-}e^{+}],\text{
eV}$ & $\Gamma\lbrack\chi_{c2}\rightarrow e^{-}e^{+}],\text{ eV}$\\\hline
$300$ & $0.060_{s}+0.009_{hs}+0.023_{h}=0.091$ & $0.036_{s}-0.040_{hs}+0.066_{h}=0.062$\\\hline
$400$ & $0.063_{s}+0.013_{hs}+0.011_{h}=0.087$ & $0.038_{s}-0.033_{hs}+0.051_{h}=0.055$\\\hline
$500$ & $0.066_{s}+0.011_{hs}+0.004_{h}=0.082$ & $0.040_{s}-0.030_{hs}+0.041_{h}=0.051$\\\hline
\end{tabular}
\label{tab:results}
\end{table}
The subscripts $s$ and $h$ denote contributions from the soft and hard photon
terms and $hs$ corresponds to the interference of these contributions. In all
cases the largest numerical contribution is provided by the ultrasoft
matrix element. This contribution is relatively large and it  weakly depends
on the factorization scale $\mu_{0}$.  Our estimates for $\Gamma\lbrack
\chi_{c1}\rightarrow e^{-}e^{+}]$ is approximately  factor 5 smaller then the estimate
in Ref.\cite{Kuhn:1979bb} and in a good agreement with the estimate in
Ref.\cite{Denig:2014fha}.    For $\Gamma\lbrack\chi
_{c2}\rightarrow e^{-}e^{+}]$  our result is five times larger than one  obtained  in
Ref.\cite{Kuhn:1979bb}. 

 From Table~\ref{tab:results} one can observe that the interference of the hard and ultrasoft contributions is  numerically large for $\chi_{c2}$ width and relatively small for $\chi_{c1}$. 
This can be explained as following.  The  imaginary part of  $h(\mu_{0})$ is numerically much larger than the real one, see Eq.(\ref{hChPT}). Further, the hard coefficient function $C^{(1)}_{\gamma\gamma}$  is real and therefore corresponding interference in the  width depends only from the real part of $h(\mu_{0})$.  The imaginary part of  $C^{(2)}_{\gamma\gamma}$  is not zero and therefore in this case the interference depends on the large imaginary part $h(\mu_{0})$ and turns out numerically large.  This observation allows one to conclude that the decay width $\chi_{c2}$ is quite sensitive to the relative sign  of parameters $R_{10}$ and $R'_{21}$.  In our estimate  we used that these parameters has the same sign, see Eq.(\ref{signR}). However if they have opposite sign then the interference contribution is negative and this  reduces  the numerical value of the  
$\Gamma\lbrack\chi_{c2}\rightarrow e^{-}e^{+}]$ by  factor 2. 

 In  Ref.\cite{Kuhn:1979bb} it was shown that  unitarity and analyticity allows one to constrain the minimal values of decay widths 
\begin{align}
\Gamma\lbrack\chi_{c1}  & \rightarrow e^{-}e^{+}]\geq\frac{3}{2}\frac{\alpha
}{k_{0}}\Gamma\lbrack J/\psi\rightarrow e^{-}e^{+}]\Gamma\lbrack\chi
_{c1}\rightarrow\gamma J/\psi]\approx0.046\text{~eV},\label{UnitB1}\\
~\Gamma\lbrack\chi_{c2}  & \rightarrow e^{-}e^{+}]\geq\left(  \sqrt
{\frac{\alpha^{2}}{9}\Gamma\lbrack\chi_{c2}\rightarrow\gamma\gamma]}%
+\sqrt{\frac{9\alpha^{2}}{20k_{0}}\Gamma\lbrack\chi_{c2}\rightarrow\gamma
J/\psi]\Gamma\lbrack J/\psi\rightarrow e^{-}e^{+}]}\right)  ^{2}%
\approx0.037\text{~eV}.\label{UnitB2}%
\end{align}
 In the presented  formalism these constrains  are always satisfied because 
 the soft contribution  has a  cut which
yields the  imaginary part required for the saturation of the bounds in
Eqs.(\ref{UnitB1}) and (\ref{UnitB2}). Therefore  all our  estimates shown in
Table I are in agreement with these inequalities. As one can see from  Table I
the hard two-photon contribution is always smaller than the limiting value in
both cases. The same observation was also made  in Ref.\cite{Kuhn:1979bb}.
This clearly  indicates   that the soft photon configuration  provides a
critically important contribution to these decay amplitudes.

The derived  approach can also be used for a description of leptonic decays of  bottonium states $\chi_{bJ}$.  
These particles have almost the same  branching fractions for  $\chi_{bJ}\rightarrow \Upsilon(1S)\gamma$ decay, see e.g. \cite{Agashe:2014kda}, 
but  at present  the widths  of these states  are not yet measured. Therefore, we cannot extract the decay coupling $f^{(b)}_{\gamma}$ using experimental data. 
Instead, we use the estimates for the corresponding widths obtained in  the model with a Cornell potential \cite{Eichten:1978tg}. 
 The corresponding values can be found in Ref.\cite{Brambilla:2004wf} and read
\begin{align}
\Gamma[\chi_{b1}]=27.8 \text{keV}, \quad  \Gamma[\chi_{b2}]=31.6 \text{keV}.  
\label{width}
\end{align} 
This gives for the dimensionless coupling in the HH$\chi$PT Lagrangian  
\begin{align}
 f^{(b)}_{\gamma}\simeq 9.4. 
  \label{fbgamma}
\end{align}
On the other hand, the width of $\Upsilon(2S)$ and branching fractions  $\Upsilon(2S)\rightarrow \chi_{bJ}\gamma$  are known \cite{Agashe:2014kda}:
\begin{align}
\Gamma
\Gamma[ \Upsilon(2S)]=32 \text{keV}, \quad Br[\Upsilon(2S)]\rightarrow \chi_{b1}\gamma]=0.06, \quad 
Br[\Upsilon(2S)]\rightarrow \chi_{b2}\gamma]=0.07.
\label{widthU2}
\end{align}
Using this values we obtain
\begin{align}
 f^{\prime(b)}_{\gamma}\simeq -16. 
 \label{fpbgamma}
\end{align}
The sign of the couplings $f^{(b)}_{\gamma}$ and $f^{\prime(b)}_{\gamma}$ in Eqs.(\ref{fbgamma})  and (\ref{fpbgamma}) is again  defined with the help of of the overlap representation as in Eq.(\ref{foverlap}) and corresponding  estimates given in  Ref.\cite{Brambilla:2004wf}. 
The corresponding radial wave functions at the origin read \cite{Eichten:1995ch}
\begin{equation}
|R_{21}^{\prime}(0)|^{2}\simeq 2.067\text{GeV}^{5},
\end{equation}%
\begin{equation}
\left\vert R_{10}(0)\right\vert ^{2}\simeq 14.05\text{GeV}^{3}, \quad \left\vert R_{20}(0)\right\vert ^{2}\simeq 5.7\text{GeV}^{3}.
\end{equation}
With these values and taking $\mu_{0}=400$MeV we obtain
\begin{align}
\Gamma\lbrack\chi_{b1}\rightarrow e^{-}e^{+}]=\left( 2.0_{s}+0.9_{hs}+1.1_{h} \right) \times 10^{-3}\simeq 4.0\times 10^{-3}\text{eV},
\\
\Gamma\lbrack\chi_{b2}\rightarrow e^{-}e^{+}]=\left(1.2_{s}-0.26_{hs}+1.6_{h} \right) \times 10^{-3}\simeq 2.6\times 10^{-3}\text{eV}.
\end{align}
We observe that in this case  the  contribution of  the ultrasoft  configuration also remains  large comparing to  the hard one. 

\section{Conclusions}
\label{conc}

The decay width $\Gamma\lbrack\chi_{cJ}\rightarrow e^{-}e^{+}]$ was computed
 using a factorization NRQCD approach.
 The dominant partonic subprocess  was described by the annihilation of the
heavy quark-antiquark pair  into two photons: $c\bar{c}\rightarrow\gamma^{\ast}\gamma^{\ast
}\rightarrow e^{+}e^{-}$ . The corresponding contribution   is given  by the
one-loop diagram with two photons in the  intermediate state. The dominant
regions in the loop integral  are associated with  two configurations: 
 hard photons and  one ultrasoft and  hard photons. 
  The soft part of the contribution with  ultrasoft
 photon overlaps with the higher Fock state $|Q\bar{Q}\gamma\rangle$ of the heavy meson, while the
 hard  contribution overlaps with the leading two quark state.    
 We have demonstrated that these  
contributions can be  factorized and described by two different operators in
NRQCD effective theory.  The ultrasoft photon contribution is estimated using
framework of the heavy hadron chiral perturbation theory. This allows us to obtain numerical
estimates using a minimal set  of the known nonperturbative parameters.  Our
estimates for charmonia $\chi_{c1}$ and $\chi_{c2}$ show that the ultrasoft photon configurations  provide the numerically 
 dominant   contribution.  This  explains why the  obtained 
numerical results for $\Gamma\lbrack\chi_{c1}\rightarrow e^{-}e^{+}]$ are  in good 
agreement with the estimates obtained  in Ref.\cite{Denig:2014fha} where only the 
usoft contribution was considered. 
 We  also expect that the developed formalism can be helpful  to
perform a  more systemic  analysis of decays if  charmonium-like  state such as  $X(3872)\rightarrow e^{-}e^{+}$. 

\section*{Aknowlegements}
We are grateful to Achim Denig for  useful discussions. This work is supported by
the Helmholtz Institute Mainz.

\end{document}